\documentclass[conference]{IEEEtran}
\IEEEoverridecommandlockouts
\usepackage{cite}
\usepackage{amsmath,amssymb,amsfonts}
\usepackage{graphicx}
\usepackage{textcomp}
\usepackage{xcolor}
\usepackage{subcaption}
\usepackage{bm}
\usepackage{multirow}
\usepackage{booktabs}
\usepackage{subcaption}
\usepackage{amsthm}\usepackage{balance}\usepackage{url}
\usepackage{algorithm,algpseudocode}
\usepackage{amssymb}
\def\BibTeX{{\rm B\kern-.05em{\sc i\kern-.025em b}\kern-.08em
    T\kern-.1667em\lower.7ex\hbox{E}\kern-.125emX}}
\begin{document}

\title{Low-Rate  Semantic Communication with Codebook-based Conditional Generative Models\\
}

\author{Kailang Ye, Mingze Gong, Shuoyao Wang, and Daquan Feng  \\
    College of Electronic and Information Engineering, Shenzhen University, China\\
    E-mail: \{yekailang2022, gongmingze2022\}@email.szu.edu.cn, \{sywang, fdquan\}@szu.edu.cn
    \thanks{This work is supported in part by the Guangdong Basic and Applied Basic Research Foundation (Project number 2025A1515010249), and in part by the Shenzhen University 2035 Program for Excellent Research (Grant No. 2024C010). Shuoyao Wang is the corresponding author. }
    }
\maketitle

\begin{abstract}
Generative semantic communication models are reshaping semantic communication frameworks by moving beyond pixel-wise optimization to align with human perception. However, many existing approaches prioritize image-level perceptual quality, often neglecting alignment with downstream tasks, which can lead to suboptimal semantic representation. 
This paper introduces an Ultra-Low Bitrate Semantic Communication (ULBSC) system that employs a conditional generative model and a learnable condition codebook.
By integrating saliency conditions and image-level semantic information, the proposed method enables high-perceptual-quality and controllable task-oriented image transmission. 
Recognizing shared patterns among objects, we propose a codebook-assisted condition transmission method, integrated with joint source-channel coding (JSCC)-based text transmission to establish ULBSC. 
The codebook serves as a knowledge base, reducing communication costs to achieve ultra-low bitrate while enhancing robustness against noise and inaccuracies in saliency detection. 
Simulation results indicate that, under ultra-low bitrate conditions with an average compression ratio of 0.57‰, the proposed system delivers superior visual quality compared to traditional JSCC techniques and achieves higher saliency similarity between the generated and source images compared to state-of-the-art generative semantic communication methods.

\end{abstract}

\begin{IEEEkeywords}
Semantic Communication, Generative Model 
\end{IEEEkeywords}

\section{Introduction}
With the rise of emerging applications like augmented reality (AR) and virtual reality (VR), the demand for efficient data transmission services has grown significantly. These services are crucial for facilitating downstream artificial intelligence (AI) tasks, ensuring they can be executed within acceptable delays. The conventional systems based on the Shannon paradigm can no longer meet the need for more efficient communication. Recently, semantic communication has emerged as a promising technique to enhance communication efficiency.

One notable system is DeepJSCC\cite{8723589}, which integrates source and channel coding within an end-to-end framework using deep learning. 
It optimizes transmission efficiency by jointly learning feature extraction and compression, rather than addressing them separately, making it highly effective for image transmission.
To further enhance the compatibility for deployment within existing communication infrastructures, DeepJSCC has been further extended to digital communication systems in DJSCC-Q\cite{9998051}. 
In addition to image transmission, several semantic communication systems have been explored for other types of data, such as text\cite{9398576} and speech\cite{9500590}. 


Although current DeepJSCC transmission systems demonstrate superior performance, they are usually optimized for mean-square error (MSE) distortion metrics, such as peak signal-to-noise ratio (PSNR), which measure the pixel-level similarity between reconstructed and original images. 
However, it is becoming increasingly evident that high pixel-level similarity does not always correlate with high perceptual quality, which better reflects the realism and visual appeal as perceived by humans. 
To enhance the perceptual quality of transmitted images, generative AI (GenAI) models have recently achieved remarkable success in enhancing communication at the semantic level.


In recent years, many studies \cite{grassucci2023generative, 10447279, 10615975} on generative semantic communication have made significant progress in selecting relevant information from the original image.
For instance, \cite{grassucci2023generative} transmitted semantic segmentation maps to generate semantically consistent scenes using diffusion models.
To improve the visual quality, \cite{10447279} proposed a system empowered by a conditional diffusion model \cite{9878449} to transmit both semantic segmentation and low-resolution compressed images for image generation at the receiver.
To further improve the interpretability of the generative semantic communication system, \cite{10615975} introduced knowledge graph and combined it with semantic segmentation maps for image generation.

Abovementioned works have achieved image generation with high semantic similarity. 
However, the transmission of semantic segmentation maps themselves results in significant communication overhead.
Thus, \cite{10446638, 10599525} considered the textual description of the source image as critical information for generative models.
To further improve semantic accuracy and visual consistence of the generated images, some works \cite{10447237, 10705025, cicchetti2024language} explored to utilize information in multiple modality for generation.
For instance, \cite{10447237} proposed a system to transmit the image caption and noisy image, which is processed based on the source, to provide precise semantics for the receiver to generate images with high semantic similarity. 
Similarly, \cite{10705025} proposed a multi-modal image transmission, where the image's caption, the position and color of key objects within the image, and the background color of the image are utilized, for efficient semantic communication. 
Although the systems\cite{10447237, 10705025, cicchetti2024language} have leveraged side information to achieve remarkable image generation performance, the data volume within them is significant due to transmission of side information, e.g. the position and color information of key objects.

\begin{figure}[!t]
    \centering
    \includegraphics[scale=0.33]{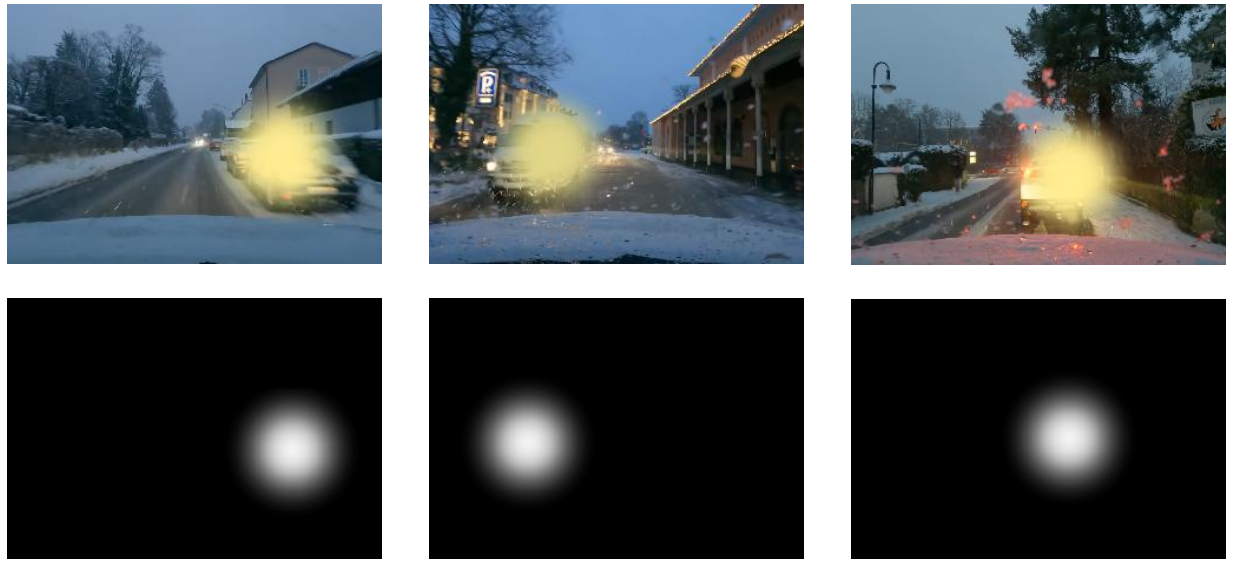}
    \caption{\small{Human eye saliency analysis in real-time driving scenarios. }}
    \label{driving scenarios}
\end{figure}

This paper studies a wireless image transmission scenario, with a particular focus on foreground objects, i.e., object saliency. For instance, in an autonomous driving scenario, it is necessary to rapidly identify hazards and respond promptly to ensure driving safety, as illustrated in Fig.~\ref{driving scenarios}. 
Instead of focusing on the background, it is more important to analyze key objects on the road, such as vehicles and pedestrians, through saliency detection. 
In particular, we propose to only transmit the position information of key objects in the image and the image caption, allowing the receiver to generate images with not only high semantic fidelity but also high image-level quality.
At the receiver, the received signal is utilized as guiding input for the GenAI model. 
To the best of the authors' knowledge, we are among the first to consider generative semantic communication for the semantic of saliency.
The contributions of this paper are summarized as follows:
\begin{itemize}
\item We propose a saliency-aware semantic communication system for low-rate transmission with codebook-based conditional generative model. 
The proposed system has advantage in generating images with both visual quality and semantic fidelity, while requiring extremely low communication overhead. 
\item We design a dual-format prompt transmission mechanism: codebook-based saliency and DJSCC-based text transmission, capitalizing on the shared patterns of saliency maps and more detail semantics in text. 
This approach enables us to reduce communication overhead and improve robustness against noise and inaccuracies in saliency extraction. 
\item The simulation results have demonstrated the superiority of the proposed method, achieving remarkable visual quality and semantic fidelity even in low signal-noise-ratio (SNR) condition. 
\end{itemize}

The remainder of this paper is organized as follows: 
In the following Section~II, we will introduce the proposed ULBSC system model.
Section~III illustrates the network architecture of the proposed semantic communication system. 
Then, simulation results are presented in Section~IV.
Finally, we will summarize this paper in Section~V.

\begin{figure*}[ht]
    \centering
    \includegraphics[scale=0.5]{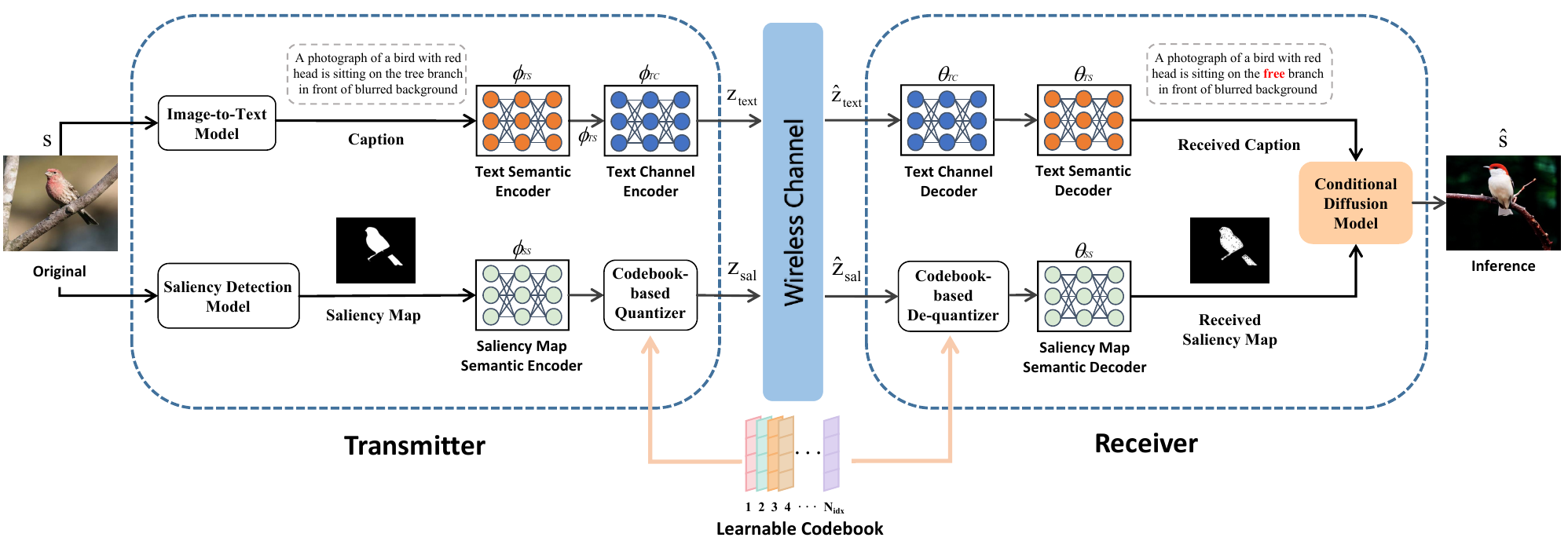}
    \caption{\small{Overview of the proposed image transmission method.}}
    \label{fig:overview}
\end{figure*}

\section{System Model}

As illustrated in Fig.~\ref{fig:overview}, the caption of the input image is extracted through image-to-text model and saliency map is extracted through pre-trained saliency detection model.
Then the extracted caption and saliency map are transmitted respectively.
The extracted caption is encoded by the pre-trained neural networks for text transmission and transmitted through the wireless channel, while the extracted saliency map is transmitted through the proposed codebook-based semantic communication approach. 
In this system, the saliency map is encoded as an index.
Then the receiver decodes the received signals and reconstructs the caption and saliency map, which are input to the pre-trained conditional diffusion model to generate a high-quality image. 

\subsection{Transmitter}
Transmitter aims at extracting semantic information from the source information to improve communication efficiency. 
As shown in Fig.~\ref{fig:overview}, the transmitter contains image-to-text model, text semantic encoder, text channel encoder, saliency detector, saliency map semantic encoder, and quantizer. 
For better understanding, the encoding process in the transmitter can be divided into two branches: the text information extraction and saliency information extraction. 
We specifically focus on exploring and designing the saliency information extraction branch, while following the method outlined in \cite{9398576} to construct the text information extraction branch.

\subsubsection{Text information extraction branch}
The input R.G.B image $\bm{s} \in \mathbb{R}^{h \times w \times 3}$, with height $h$ and width $w$, is processed by an image-to-text model to generate a descriptive caption for the input image.
Notably, converting images into descriptive text can greatly reduce the communication overhead while making information of visual content more accessible.

The image-to-text model summarizes $\bm{s}$ into a sentence $\bm{w}=[w_1, w_2, ..., w_\mathrm{L}]$, where $w_l$ represents the $l$-th word in the sentence. 
The image-to-text process can be given by 
\begin{equation}
    \bm{w} = E_\mathrm{i2t}(\bm{s}), 
\end{equation}
where $E_\mathrm{i2t} :: \mathbb{R}^{h \times w \times 3} \to \mathbb{R}^{L}$ is the process function of image-to-text model. 
In addition, $L$ is the length of the output sentence, which is also called caption in this paper.
Then, the caption is sequentially processed by a text semantic encoder to further extract the semantic information and a text channel encoder to compress and enhance the transmitted signal.
The encoded symbol stream of the text information extraction branch can be represented by 
\begin{equation}
    \bm{z}_\mathrm{text} = E_\mathrm{TC}(E_\mathrm{TS}(\bm{w}, \bm{\phi}_\mathrm{TC}), \bm{\phi}_\mathrm{TS}), 
\end{equation}
where $E_\mathrm{TS} :: \mathbb{R}^{L} \to \mathbb{R}^{L \times V}$ is the semantic encoding function with learnable parameters $\bm{\phi}_\mathrm{TS}$. 
Moreover, $E_\mathrm{TC} :: \mathbb{R}^{L \times V} \to \mathbb{R}^{N}$ is the channel encoding function with learnable parameters $\bm{\phi}_\mathrm{TC}$ for text transmission. 
Additionally, $V$ is the dimension of encoding neural network's output and $N$ is the number of symbols in the transmitted signal. 


\subsubsection{Saliency information extraction branch}
Different from previous approaches in semantic communication for R.G.B image transmission\cite{8723589}, we transmit the gray-scale saliency map within the proposed system. 
The input image $\bm{s}$ is first processed by the saliency detection model, whose output is a saliency map $\bm{m}$. 
It is worth noting that a saliency map is a representation highlighting the important or noticeable regions in an image. 
Moreover, it provides insights into the visualization application in the receiver side, while reducing the communication overhead significantly.

Given the input image $\bm{s}$, the saliency detection model detects key objects and outputs a gray-scale saliency map $\bm{m} \in \mathbb{R}^{h \times w \times 1}$. 
The detection process is empowered by a learned function $E_\mathrm{i2s} :: \mathbb{R}^{h \times w \times 3} \to \mathbb{R}^{h \times w \times 1}$. 
Then, the saliency map is compressed by the saliency map semantic encoder through a learned function $E_\mathrm{SS} :: \mathbb{R}^{h \times w \times 1} \to \mathbb{R}^{h' \times w' \times c'}$, where $h'$, $w'$, and $c'$ are the dimensions of the last layer's output. 
Inspired by \cite{NIPS2017_7a98af17}, a codebook-based quantizer is applied following the saliency map-based semantic encoding, effectively reducing redundancy in the extracted semantic features and enhancing communication efficiency.
Briefly, the encoded symbol stream of the saliency information extraction branch is given by 
\begin{equation}
    \bm{z}_\mathrm{sal} = Q(E_\mathrm{SS}(\bm{m}, \bm{\phi}_\mathrm{SS}), \bm{\phi}_\mathrm{codebook}), 
\end{equation}
where $Q :: \mathbb{R}^{h' \times w' \times c'} \to \mathbb{R}^{1}$ is the codebook-based quantization function with learnable parameters $\bm{\phi}_\mathrm{codebook}$ and $\bm{\phi}_\mathrm{SS}$ is the learnable parameter of saliency map semantic encoding function $E_\mathrm{SS}$. 
After extracting both types of semantic features, the transmitter normalizes the signal power of the data and sends it through the wireless channel for transmission.


\subsection{Physical Channel}
Assuming two independent communication link for the text and saliency map transmission, to simulate the process of physical information transmission, we have employed the additive white Gaussian noise (AWGN) channel model widely used in wireless communication as the physical communication channel. 
Both $\bm{z}_\mathrm{text}$ and $\bm{z}_\mathrm{sal}$ undergo transmission through the physcial channel to reach the receiver. 
The signals received at the receiver are denoted as $\bm{\hat{z}}_\mathrm{text}$ and $\bm{\hat{z}}_\mathrm{sal}$, respectively. 
In brief, the process can be represented as follows:
\begin{equation}
    \begin{bmatrix}
        \hat{\bm{z}}_{\text{text}} \\
        \hat{\bm{z}}_{\text{sal}}
    \end{bmatrix}
    =
    \begin{bmatrix}
        \bm{z}_{\text{text}} \\
        \bm{z}_{\text{sal}}
    \end{bmatrix}
    + \bm{n}
\end{equation}
where $n$ follows the Gaussian distribution $\mathcal{N}(0, \sigma^2 \bm{I})$.



\subsection{Receiver}
As the semantic features are received, the receiver leverages the received information and the image generation model to display images whose visual content are highly similar with that in the source image.
Notably, the receiver contains a text channel decoder, a text semantic decoder, a de-quantizer, a saliency map semantic decoder, and a conditional diffusion model. 
In particular, the text channel decoder takes the received text semantic feature $\hat{\bm{z}}_\mathrm{text}$ as input to mitigate the negative impact of noise corruption through a learned function $D_\mathrm{TC} :: \mathbb{R}^{N} \to \mathbb{R}^{L \times V}$.
Following the text channel decoder, a text semantic decoder is deployed to reconstruct the source caption through a learned function $D_\mathrm{TS} :: \mathbb{R}^{L \times V} \to \mathbb{R}^{L}$.
Therefore, the text decoding process can be given by 
\begin{equation}
    \hat{\bm{w}} = D_\mathrm{TS}(D_\mathrm{TC}(\bm{\hat{z}}_\mathrm{text}, \bm{\theta}_\mathrm{TC}), \bm{\theta}_\mathrm{TS}), 
\end{equation}
where $\bm{\theta}_\mathrm{TC}$ and $\bm{\theta}_\mathrm{TS}$ are the trainable parameters. 
Similarly, as for the received saliency semantic feature, the decoding process can be represented by 
\begin{equation}
    \hat{\bm{m}} = D_\mathrm{SS}(DQ(\bm{\hat{z}}_\mathrm{sal}, \bm{\phi}_\mathrm{codebook}), \bm{\theta}_\mathrm{SS}), 
\end{equation}
where $DQ$ is the de-quantization function and $D_\mathrm{SS} :: \mathbb{R}^{h' \times w' \times c'} \to \mathbb{R}^{h \times w \times 1}$ is the saliency map semantic decoding function with learnable parameter $\bm{\theta}_\mathrm{SS}$. 
As the decoding processes are finished, both of $\hat{\bm{w}}$ and $\hat{\bm{m}}$ are input into the conditional diffusion model to generate user-specific images. 

Recently, diffusion model has demonstrates impressive performance in improving efficiency in semantic communication system. 
Building upon diffusion model, conditional diffusion model leverages the additional information, such as saliency map, to allow the users to generate specific data, while maintaining high fidelity in contents. 
Therefore, in this paper, we employ a pre-trained conditional diffusion model in the receiver to display images with high visual quality and semantic similarity. 
Briefly, the inference process can be given by 
\begin{equation}
    \hat{\bm{s}} = CD(\hat{\bm{w}}, \hat{\bm{m}}), 
\end{equation}
where $\hat{\bm{s}}$ is the generated image and $CD$ is the conditional diffusion function.


\section{Network Architecture}

\subsection{Image-to-Text Model}
To extract text representation from the source image, we use bootstrapping language-image pre-training for unified vision-language understanding and generation (BLIP)\cite{li2022blip} for image captioning. 
BLIP combines an encoder and decoder together during pre-training, allowing for multi-modal space alignment for both understanding and generation objectives. 
BLIP is composed of three parts: an encoder for extracting image and text features, an image-grounded text encoder whose inputs are image and text features, and an image-grounded text decoder, which outputs text based on input image features.

\subsection{Text Semantic Codec and Text Channel Codec}
As mentioned in Section~I, DeepSC\cite{9398576} has demonstrated impressive efficiency in semantic communication for text transmission. 
Inspired by its advantages, we adopt DeepSC as the semantic communication model, including text semantic codec and text channel codec, to transmit the extracted image caption. 
We follow \cite{9398576} to build our model for text transmission. 
In particular, both of text semantic encoder and decoder are constructed based on Transformer blocks \cite{vaswani2017attention}. 
Additionally, text channel encoder and decoder consist of linear layers.



Following \cite{9398576}, the loss function for training text transmission model consists of $\mathcal{L}_\mathrm{CE}$ and $\mathcal{L}_\mathrm{MI}$. 
$\mathcal{L}_\mathrm{CE}$ is the cross entropy loss, aiming to minimize the semantic difference between the source caption $\bm{w}$ and the reconstructed caption $\hat{\bm{w}}$. 
$\mathcal{L}_\mathrm{CE}$ is given by 
\begin{align}
&\mathcal{L}_\mathrm{CE}(w, \hat{w}; \bm{\phi}_\mathrm{TS}, \bm{\phi}_\mathrm{TC}, \bm{\theta}_\mathrm{TS}, \bm{\theta}_\mathrm{TC})  \notag \\
&= -\sum_{l=1} \left( q(w_l) \log(p(w_l)) + (1 - q(w_l)) \log(1 - p(w_l)) \right), 
\end{align}
where \( q(w_l) \) is the real probability that the \( l \)-th word, \( w_l \), appears in the estimated sentence \( \bm{w} \), and \( p(w_l) \) is the predicted probability that the \( i \)-th word, \( w_i \), appears in the sentence \( \hat{\bm{w}} \).

Furthermore, $\mathcal{L}_\mathrm{MI}$ is the mutual information between the transmitted text semantic feature $\bm{z}_\mathrm{text}$ and the received one $\hat{\bm{z}}_\mathrm{text}$, aiming to maximize the achieved data rate during the training of transmitter. 
$\mathcal{L}_\mathrm{MI}$ can be given by 
\begin{equation}
\mathcal{L}_{MI}(\bm{z}, \hat{\bm{z}}; \bm{\phi}_\mathrm{TS}, \bm{\phi}_\mathrm{TC}) 
= \mathbb{E}_{p(\bm{z}, \hat{\bm{z}})}[f_T] - \log \left( \mathbb{E}_{p(\bm{z})p(\hat{\bm{z}})} \left[ e^{f_T} \right] \right),
\end{equation}
where $T$ is the neural network used for estimating the mutual information\cite{9398576} and $f_T$ is the process function of it. 
The total loss function for training text transmission model  is given by 
\begin{align}
    &\mathcal{L}_\mathrm{text} (\bm{\phi}_\mathrm{TS}, \bm{\phi}_\mathrm{TC}, \bm{\theta}_\mathrm{TS},\bm{\theta}_\mathrm{TC}) = \mathcal{L}_{\text{CE}} - \eta \mathcal{L}_\mathrm{MI},
\end{align}
where $\eta (0 \leq \eta \leq 1)$ is a hyper parameter to balance $\mathcal{L}_\mathrm{CE}$ and $\mathcal{L}_\mathrm{MI}$. 
In this paper, we follow the setting of $\eta$ in \cite{9398576}. 

\begin{figure}[!t]
    \centering
    \includegraphics[scale=0.45]{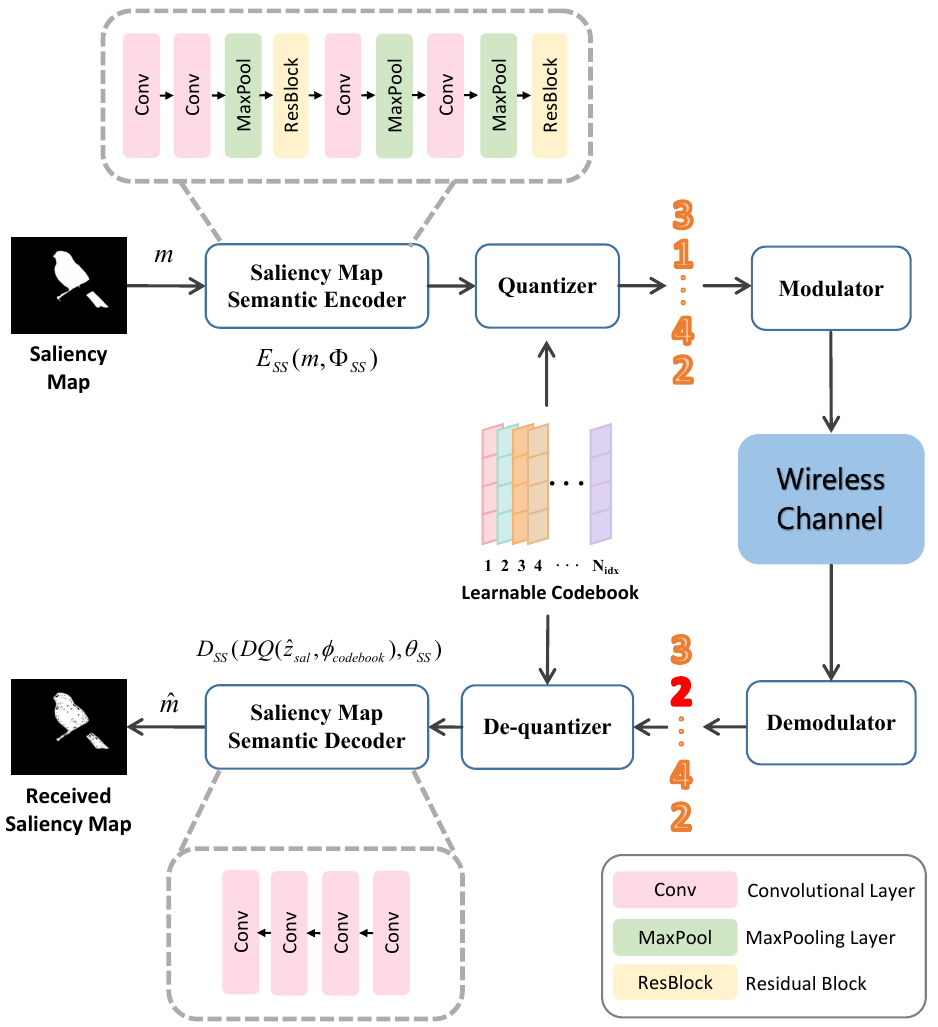}
    \caption{\small{The proposed model for saliency map as the condition.}}
    \label{fig:my_label}
\end{figure}

\subsection{Saliency Detection Model}

To accurately detect the saliency of key objects in the source image, we adopt a pre-trained salient object detection model\cite{8315520}. 
The adopted model mainly consists of covolutional layers. 
It is worth noting that a lightweight version of the model\cite{8315520} is selected for efficient extraction of saliency map. 

\subsection{Saliency Map Semantic Codec and Codebook-Based Quantizer}

The side information in saliency map extracted in the transmitter is critical for the receiver to generate high visual-perceptual quality images.
Therefore, as shown in Fig.~\ref{fig:my_label}, we develop saliency map codec and a codebook-based quantizer to achieve efficient transmission of saliency map. 
Inspired by previous success in image transmission-oriented semantic communication systems\cite{8723589}, we build both of the saliency map semantic encoder and decoder with convolutional layers. 
Furthermore, we introduce residual block (ResBlock) \cite{he2016deep} to construct the encoder and decoder to enhance the semantic accuracy and noise robustness of the latent representation. 

The transmission of gray-scale saliency map can be further improved beyond previous framework in \cite{8723589}. 
It can be observed that only part of the gray-scale map is highlighted for the key objects. 
In other words, the information density of the map is much lower than that of regular images, whose transmission is considered in \cite{8723589}. 
To further remove the redundancy in the map to improve transmission efficiency, we propose a codebook-based quantizer with learnable parameters $\bm{C}=[\bm{c}_1, \bm{c}_2, ..., \bm{c}_{N_\mathrm{idx}}] \in \mathbb{R}^{D \times N_\mathrm{idx}}$, where $D$ is the code-length of each code in the proposed codebook,to convert the extracted map features into indices. 
Notably, the codebook is shared by the transceivers. 
Furthermore, all of saliency semantic encoder, decoder, and the codebook are jointly trained in an end-to-end manner. 
Thus, the loss function for the training of saliency map transmission can be given by 
\begin{equation}
    \mathcal{L}_\mathrm{sal}(\bm{\phi}_\mathrm{SS}, \bm{\theta}_\mathrm{SS}, \bm{C}) = \frac{1}{N} \sum_{i=1}^{N} (\bm{m}_i - \hat{\bm{m}}_i)^2 - \lambda \cdot \mathbb{E}[\log p(z_\text{sal})], 
\end{equation}
where $\bm{m}_i$ is the original saliency map and $\hat{\bm{m}}_i$ is the generated saliency map. 
Moreover, $p(z_\text{sal})$ is the probability distribution of the latent variable $z_\text{sal}$, typically a standard normal distribution.
In addition, $\lambda (0 \leq \lambda \leq 1)$ is the hyper parameter for balancing two terms in $\mathcal{L}_\mathrm{sal}$. 
In this paper, we set $\lambda$ to $0.001$. 

\subsection{Conditional Diffusion Model}

As for the conditional diffusion model part in the receiver, we adopt Stable Diffusion\cite{9878449} for image generation. 
Stable Diffusion models have demonstrated remarkable generation ability in computer vision field and the introduction of side information is allowed to satisfied user-specific image generation. 
According to user requirement in images with high visual-perceptual and semantic similarity, we adopt an open-source pre-trained model \footnote{https://huggingface.co/stable-diffusion-v1-5/stable-diffusion-v1-5} to generate image based on saliency map and image caption.

\section{Simulation Results}


\subsection{Simulation Settings}
The DUTS dataset comprises two parts: DUTS-TR and DUTS-TE, containing 10,553 training images and 5,019 test images, respectively.
All training images are sourced from the ImageNet DET training and validation sets, while the test images are obtained from the ImageNet DET test set and the SUN dataset.
Both the training and test sets present highly challenging scenarios for saliency detection, with pixel-level ground truth manually annotated by 50 subjects.
Therefore, DUTS-TR was selected as the training set, while DUTS-TE was chosen as the test set.
To transmit captions, we utilize the JSCC framework with the NNs in DeepSC.
For the training data, we used the pre-trained CLIP model to generate a corpus of 15,000 captions, with all captions derived from the DUTS dataset.
This corpus is then used to train DeepSC under varying SNR conditions from -3 dB to 15 dB and the trained model is subsequently saved.
To transmit saliency map, we train the communication models with the DUTS-TE dataset.




\subsection{Simulation Metric}
\subsubsection{F-measure}
The precision value is the ratio of ground truth salient pixels in the predicted salient region. And the recall value is defined as the percentage of the detected salient pixels and all ground truth area. The precision and recall are calculated by thresholding the predicted saliency map and comparing it with the corresponding ground truth.$F_{\beta}$ is used to comprehensively evaluate both precision and recall as:
\begin{equation}
    F_\beta = \frac{(1 + \beta^2) \times \text{Precision} \times \text{Recall}}{\beta^2 \times \text{Precision} + \text{Recall}}, 
\end{equation}
where the hyper parameter $\beta^2$ is set to 0.3, following the previous work \cite{8578285}.

\subsubsection{Mean Absolute Error (MAE)}
MAE denotes the average per-pixel difference between a predicted saliency map and its ground truth mask. It is given by:
\begin{equation}
\text{MAE} = \frac{1}{h \times w} \sum_{r=1}^{h} \sum_{c=1}^{w} | P(r, c) - G(r, c) |, 
\end{equation}
where $P$ and $G$ are the probability map of the salient object detection and the corresponding ground truth, respectively.
Furthermore, $(r, c)$ are the pixel coordinates.

\begin{figure}[!t]
    \centering
    \subfloat[MAE versus SNR]{
        \includegraphics[width=0.45\linewidth]{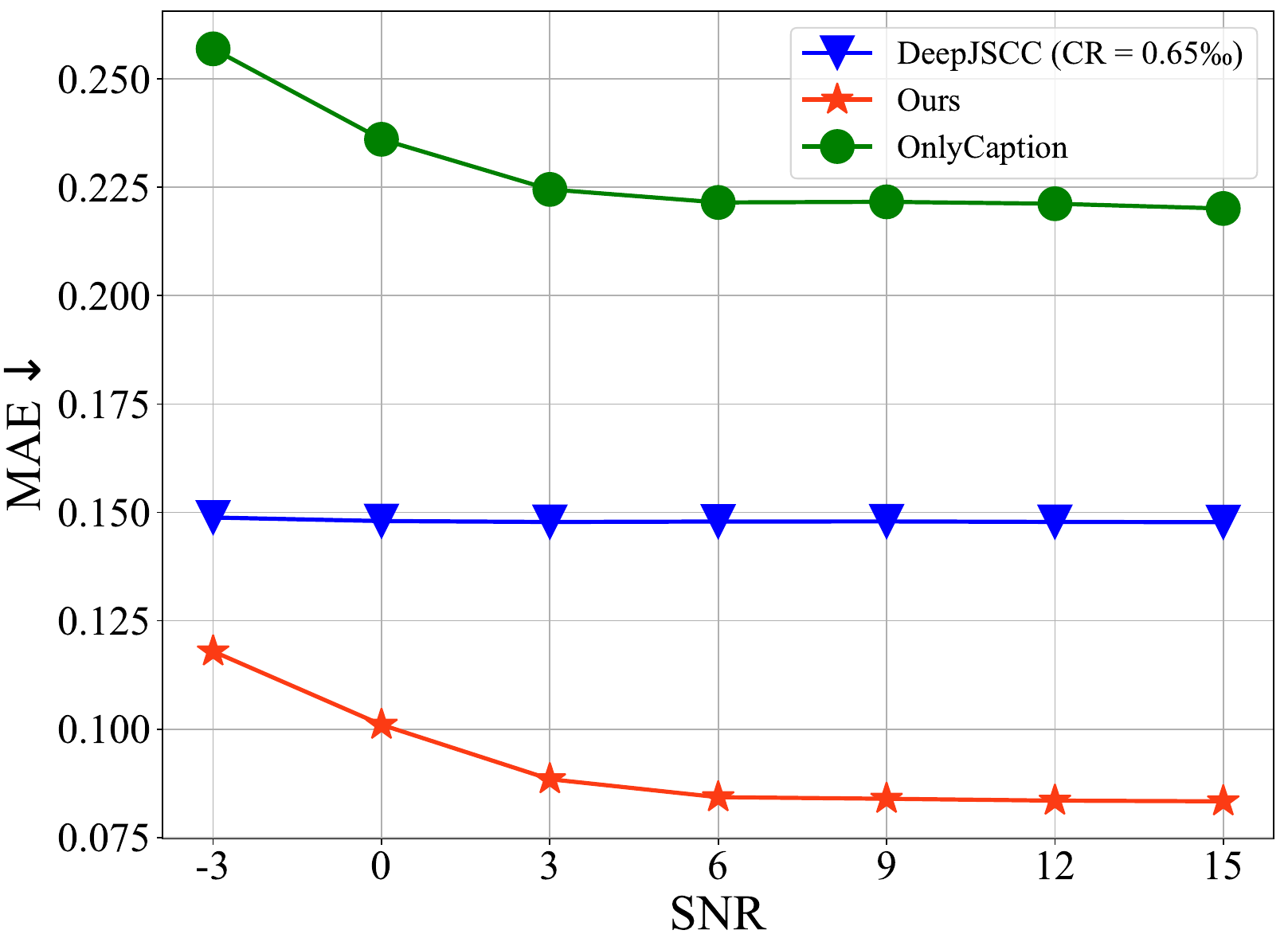}%
        \label{fig:mae}}
    \subfloat[F-measure versus SNR]{
        \includegraphics[width=0.44\linewidth]{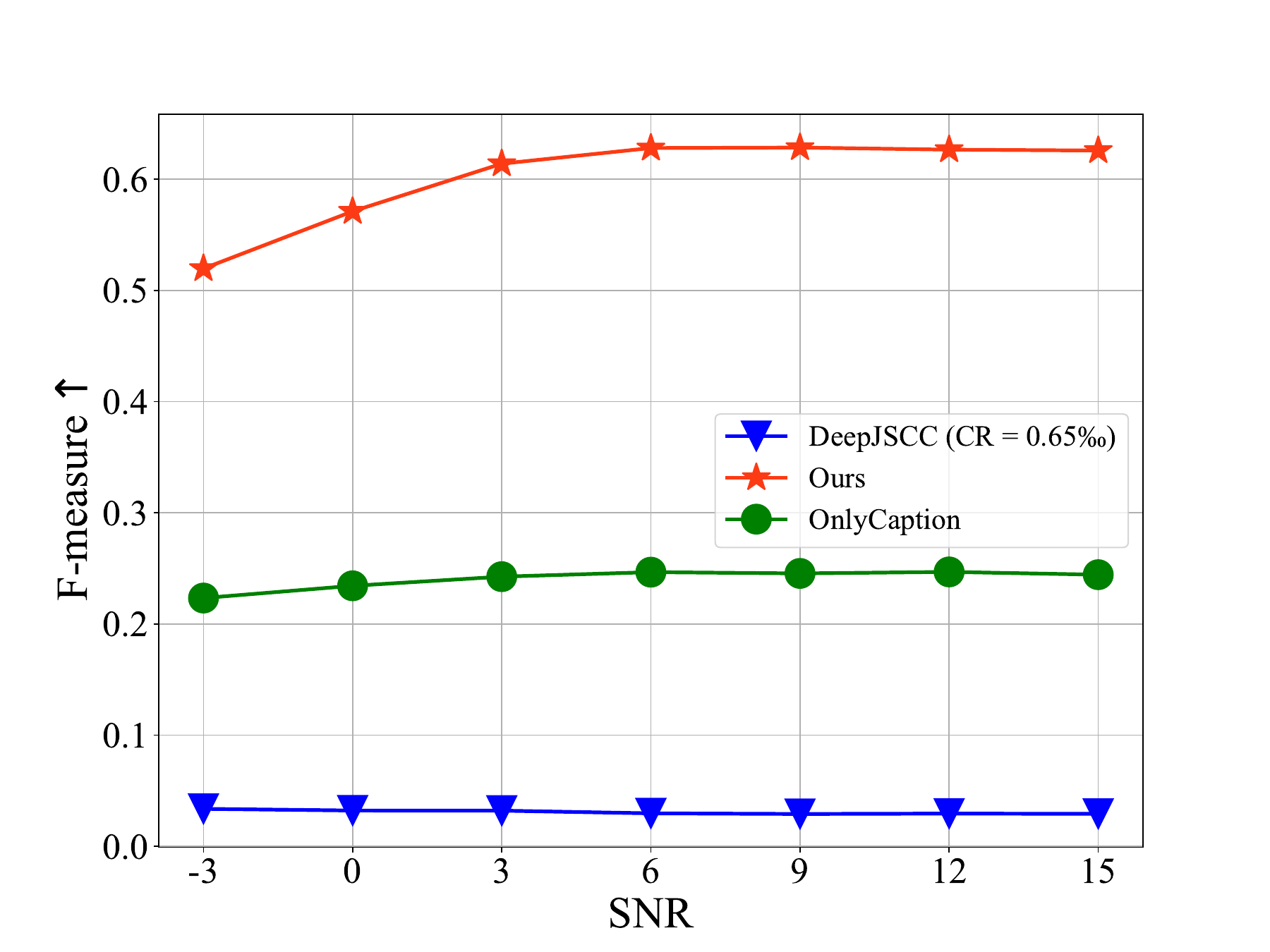}%
        \label{fig:f_measure}}
    \caption{\small{Saliency detection accuracy versus SNR. }}
    \label{fig:curves}
\end{figure}

\begin{table}[!t]
  \centering
  \caption{\small{The communication overhead (KB) for generative image transmission. In \cite{10705025}, the average compression ratio is 50‰, whereas the average compression ratio is 0.57‰ in our work.
}}
    \resizebox{1\linewidth}{!}{
    \begin{tabular}{|c|c|c|c|c|}
    \hline
    Original & Saliency Map & Caption\&Embedding\cite{cicchetti2024language}&Caption & Ours \\
    \hline
    24.548 & 0.311 & 0.525&\textbf{0.014} & \textbf{0.014} \\
    \hline
    \end{tabular}%
  }
  \label{tab:addlabel}%
\end{table}%

\begin{figure*}[!t]
    \centering
    \includegraphics[clip, trim={0 0.3cm 0 0.05cm}, scale=0.68]{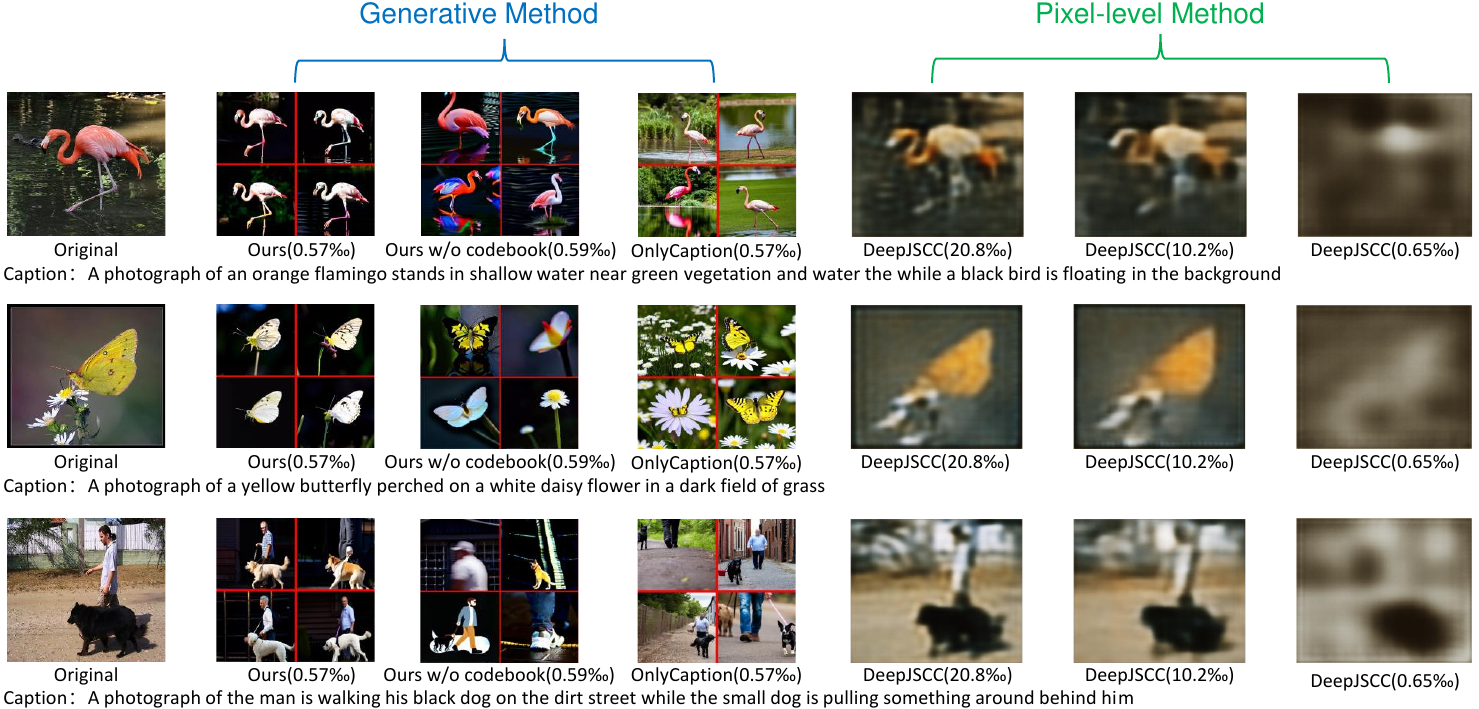}
    \caption{\small{Visual comparison of image transmission at a test SNR of 12 dB in the AWGN channel. For ``Ours w/o Codebook", the codebook is not used to assist in transmitting the saliency map. Instead, we utilize the DeepJSCC\cite{8723589} method for transmission.}}
    \label{fig:example}
\end{figure*}

\subsection{Benchmarks}
To demonstrate the validity of our proposed ULBSC, we chose two benchmarks:
\begin{itemize}
\item \textbf{DeepJSCC\cite{8723589}}: This benchmark is initially proposed for pixel-oriented communications, where map the data to the channel input symbols by DNN-based encoders.
\item \textbf{OnlyCaption\cite{10446638}}: This benchmark is an open source training-free model based on CLIP and stable diffusion. In this benchmark, we extract the caption of the original image with CLIP and directly feed it as the input of stable diffusion to get the generated image. 
\end{itemize}

\subsection{Validation Results}

Fig.~\ref{fig:curves} illustrates the variation in two metrics of saliency detection, including MAE and F-measure, for the three semantic communication approaches versus a wide range of SNR in the AWGN channel. 
Overall, compared with two benchmarks, the proposed ULBSC achieves the best saliency detection accuracy in both metrics.
In particular, as shown in Fig.~\ref{fig:mae}, ULBSC displays 59.20\% and 37.98\% less MAE than that achieved by OnlyCaption and DeepJSCC, respectively. 
Moreover, as shown in Fig.~\ref{fig:f_measure}, ULBSC demonstrates 150.1\% and 1854\% higher F-measure values than those achieved by the others, respectively. 
These improvements underscore the effectiveness of the proposed ULBSC. 

In addition, Table.~\ref{tab:addlabel} shows that the communication overhead of our ULBSC and OnlyCaption is nearly identical. 
In other words, the overhead for saliency map transmission is succesfully reduced by the proposed codebook-based quantizer. 
This observation further supports ULBSC's superiority. 

To further demonstrate the effectiveness of ULBSC, we provide a set of visually intuitive results in Fig.~\ref{fig:example}. 
The visual comparisons are conducted under AWGN channel with noise intensity of 12dB. 
As shown in Fig.~\ref{fig:example}, the images generated by ULBSC and OnlyCaption have key objects similar with those in the original images. 
However, it can be observed that the images generated by ULBSC have higher saliency similarity with the original images than those generated by the OnlyCaption. 
Furthermore, under the similar constraint on communication overhead, ULBSC generates images with much higher visual perceptual-quality than DeepJSCC.

\section{Conclusion}
In this paper, we propose an ultra-low bitrate semantic communication (ULBSC) system, which is empowered by a conditional generative model and a learnable condition codebook. 
To overcome the misalignment with downstream tasks, the proposed method integrates saliency conditions and image-level semantic formation to enable high-perceptual quality and task-oriented image transmission. 
Based on the observation that saliency and objective features shared some patterns, we combine the codebook-assisted conditional transmission and JSCC-based text transmission to reduce the communication oveahead. 
From the perspective of saliency detection accuracy, our ULBSC achieves 59.20\% less MAE and 150.1\% higher F-measure than those achieved by OnlyCaption.

\bibliographystyle{ieeetr}  
\bibliography{references}  

\end{document}